\documentclass[pmlr]{jmlr}

\usepackage{longtable}

\usepackage{booktabs}
\usepackage[load-configurations=version-1]{siunitx} 

\theorembodyfont{\upshape}
\theoremheaderfont{\scshape}
\theorempostheader{:}
\theoremsep{\newline}

\jmlrvolume{1}
\jmlryear{2024}
\jmlrworkshop{FM-EduAssess at NeurIPS 2024 Workshop}

\title[Towards Scalable Automated Grading]{Towards Scalable Automated Grading: Leveraging Large Language Models for Conceptual Question Evaluation in Engineering}

\author{
\\[0.2cm] 
\Name{Rujun Gao\textsuperscript{1}} \Email{grj1214@tamu.edu}\\
\Name{Xiaosu Guo\textsuperscript{2}} \Email{xxg230002@utdallas.edu}\\
\Name{Xiaodi Li\textsuperscript{4}} \Email{li.xiaodi@mayo.edu}\\
\Name{Arun Balajiee Lekshmi Narayanan\textsuperscript{3}} \Email{arl122@pitt.edu}\\
\Name{Naveen Thomas\textsuperscript{1}} \Email{naveenthomas@tamu.edu}\\ 
\Name{Arun R. Srinivasa\textsuperscript{1}} \Email{arun-r-srinivasa@tamu.edu}\\
\\[0.3cm] 
\textsuperscript{1} \addr J. Mike Walker ’66 Department of Mechanical Engineering, Texas A\&M University \\
\textsuperscript{2} \addr Computer Science Department, University of Texas at Dallas \\
\textsuperscript{3} \addr Intelligent Systems Program University of Pittsburgh \\
\textsuperscript{4} \addr Department of Artificial Intelligence and Informatics, Mayo Clinic \\
}

\begin{document}

\maketitle
\vspace{-1cm}

\begin{abstract}
This study explores the feasibility of using large language models (LLMs), specifically GPT-4o (ChatGPT), for automated grading of conceptual questions in an undergraduate Mechanical Engineering course. We compared the grading performance of GPT-4o with that of human teaching assistants (TAs) on ten quiz problems from the MEEN 361 course at Texas A\&M University, each answered by approximately 225 students. Both the LLM and TAs followed the same instructor-provided rubric to ensure grading consistency. We evaluated performance using Spearman's rank correlation coefficient and Root Mean Square Error (RMSE) to assess the alignment between rankings and the accuracy of scores assigned by GPT-4o and TAs under zero- and few-shot grading settings. In the zero-shot setting, GPT-4o demonstrated a strong correlation with TA grading, with Spearman’s rank correlation coefficient exceeding 0.6 in seven out of ten datasets and reaching a high of 0.9387. Our analysis reveals that GPT-4o performs well when grading criteria are straightforward but struggles with nuanced answers, particularly those involving synonyms not present in the rubric. The model also tends to grade more stringently in ambiguous cases compared to human TAs. Overall, ChatGPT shows promise as a tool for grading conceptual questions, offering scalability and consistency.
\end{abstract}

\section{Introduction}
\label{sec:intro}

Large Language Models (LLMs) have become powerful tools capable of understanding and processing natural human language, offering significant applications in various fields, including education. In educational contexts, LLMs have been employed to reduce the grading workload by automating the assessment of student responses, particularly in large classroom settings~\citep{bonner2023large}. In STEM subjects, a deep understanding of academic concepts is critical for student success. Conceptual questions, which often require constructed short-answer responses, provide valuable insights into students' understanding from both a breadth and depth perspective. However, grading such responses is challenging due to the inherent variability in answers and the subjective nature of grading by different instructors or teaching assistants (TAs).

This process is also time-consuming and requires a substantial investment of time and expertise~\citep{kuechler2010performance}. This is particularly challenging in STEM disciplines like mechanical engineering where part of the learning is the use of precise technical language to communicate specific concepts.

Existing research on the automatic assessment of short-answer questions spans a variety of approaches~\citep{GAO2024100206}, including natural language processing (NLP), machine learning, concept mapping, and, more recently, the application of models like ChatGPT~\citep{chang2024automatic, burrows2015eras, bonthu2021automated, putnikovic2023embeddings}. The integration of LLMs, such as ChatGPT, into educational assessment has garnered significant attention due to their ability to interpret and evaluate complex text-based responses with greater consistency and scalability.

Automated grading systems have evolved from simple rule-based algorithms to more advanced models that leverage transformer-based architectures, such as BERT~\citep{Devlin2019BERTPO} and GPT. These models have significantly improved the accuracy and reliability of automated grading systems, particularly in the assessment of open-ended questions such as essays and short answers. The automatic grading system offers several advantages, particularly in large classroom settings where managing a large volume of assessments can be overwhelming: 1) it enables more consistent grading and timely feedback; 2) it allows educators to shift their focus from grading to more instructional activities and personalized student support.

While most cited works focus predominantly on automated grading in scientific fields such as biology and computer science, there is a notable gap in the literature regarding its application in other engineering disciplines, including civil, mechanical, electrical, and chemical engineering. This study addresses this gap by exploring the feasibility of using ChatGPT, specifically the GPT-4o model, to grade conceptual questions within Mechanical Engineering (ME). Our experiment employs 10 quizzes and associated graded datasets from an undergraduate ME course in materials and manufacturing at Texas A\&M University, a course rich in conceptual and technical language. To evaluate GPT-4o’s grading performance, we use Spearman’s rank correlation coefficient to compare the alignment between student rankings from the LLM-based grading and human grading, and the root mean square error (RMSE) to measure score discrepancies. Additionally, visualizations of the grading outcomes are included to provide an intuitive representation of the model's performance.

This study is motivated by the need to enhance student comprehension of engineering concepts through rapid, reliable feedback aligned with instructor rubrics in large classes. In a typical engineering class with over 100 students, providing timely and consistent feedback is notoriously challenging—yet timely feedback is crucial for effective learning (see, e.g., \citet{barboza2016importance}). An AI-assisted grading and feedback approach enables a student-centered, formative grading process \citep{henderson2019conditions}, where students engage interactively to deepen their understanding by revising answers and receiving re-evaluation \citep{GAO2024100206}. This approach, termed "supervised practice with feedback," is a cornerstone of mastery-based personalized learning, yet it is practically unfeasible in large engineering classes with hundreds of students. As highlighted in a systematic review on the benefits of automated grading by \citet{hahn2021systematic}, automated grading and feedback offer several learning advantages, including consistency, reduced grading bias, and increased student participation.

 The study aims to address two key research questions: (a) How effectively can ChatGPT be used to grade Mechanical Engineering conceptual questions? and (b) How does the in-context learning approach affect ChatGPT's grading performance?

\section{Prior Work}
\label{related_work}
In recent years, several automated methods have been explored for scoring and grading short answers. For instance, \citet{yaneva2023extracting} investigated the use of BERT with additional features to evaluate clinical essays in the medical field. Other studies have explored BERT-based joint learning models for essay scoring and feedback generation \citep{wang2022use}, facilitating large-scale evaluation of short answers, albeit with limited personalization. Distance metrics have also been employed in methods that assess and score essays automatically \citep{clark-etal-2019-sentence}. Additionally, \citet{gao2023work} compared seven open-source LLM models for automated grading of text-based short-answer questions using correct/incorrect labels.

Beyond scoring, researchers have increasingly focused on feedback generation. For example, \citet{lu2021integrating} examined various methods for incorporating feedback. Studies have demonstrated the impact and utility of feedback derived from automated essay evaluation systems \citep{liu2016automated}. Recent research has also investigated automated assessment in interactive LLM-student environments \citep{han2023llm}, while other studies have explored the application of argument mining approaches \citep{nguyen2018argument} for evaluation and scoring.

An alternative approach to evaluating student answers recently explored the use of prompt engineering, enabling students to demonstrate a certain level of understanding~\citep{10.1145/3657604.3662039}. Other approaches focus on categorizing these responses~\citep{schneider2023towards}. ~\citet{ivanova2024evaluating} conducted a comparative study with human annotators to evaluate the effectiveness of ChatGPT’s automatic grading of student answers. ~\citet{lagakis2024evaai} proposed a multi-agent framework that integrates a reviewer/grader working alongside an LLM evaluator to achieve more accurate automatic grading results.

Building on prior research in automated grading, which has been applied to various short-answer assessments in fields such as medicine and language learning, this paper examines the application of automated grading in mechanical engineering. Utilizing the latest models, this study provides multi-scale scoring aligned with instructor-provided rubrics to support and enhance student learning in engineering contexts.

\section{Methodology}
\label{method}
To investigate the extent to which large language models (LLMs) can be utilized for grading Mechanical Engineering conceptual questions, the methodology is structured as shown in figure 1. Human grading by teaching assistants (TA) serves as the benchmark. The TA grading process involves one TA and one grader, with their assessment trained and verified by the course’s instructor based on specific grading rubrics that was created by the instructor.

To evaluate the performance of LLM-based grading, as outlined in research question one, we utlized the state-of-the-art ChatGPT model (GPT-4o, released on May 13, 2024). The prompts for automated grading were designed to align with the same rubric used by the TA.

Additionally, to address the second research question, we conducted two types of experiments: 1) GPT-4o zero-shot grading and 2) GPT-4o few-shot grading, to explore how in-context learning impacts the model's grading performance. For the few-shot experiments, we provided four grading example responses, along with the TA's corresponding scores, to the GPT model. The specific prompt details are included in \appendixref{appendix_b}.

\begin{figure}[htbp]
\floatconts
  {fig:method_workflow}
  {\caption{Flow of Evaluation}}
  {\includegraphics[width=0.8\linewidth]{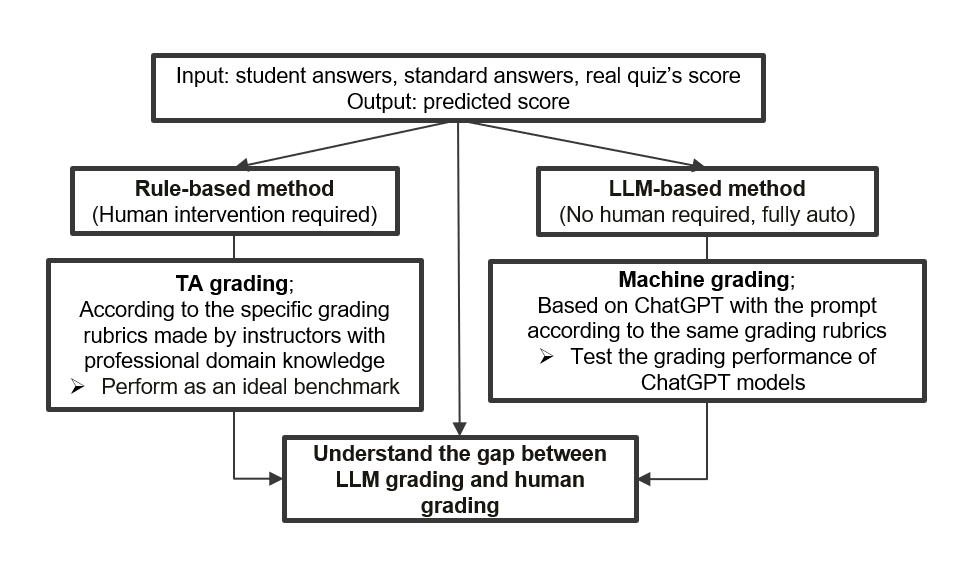}}
\end{figure}

\section{Experimental Datasets}
\subsection{Conceptual Question Example}

\begin{figure}[htbp]
    \centering 
    \includegraphics[width=0.8\linewidth]{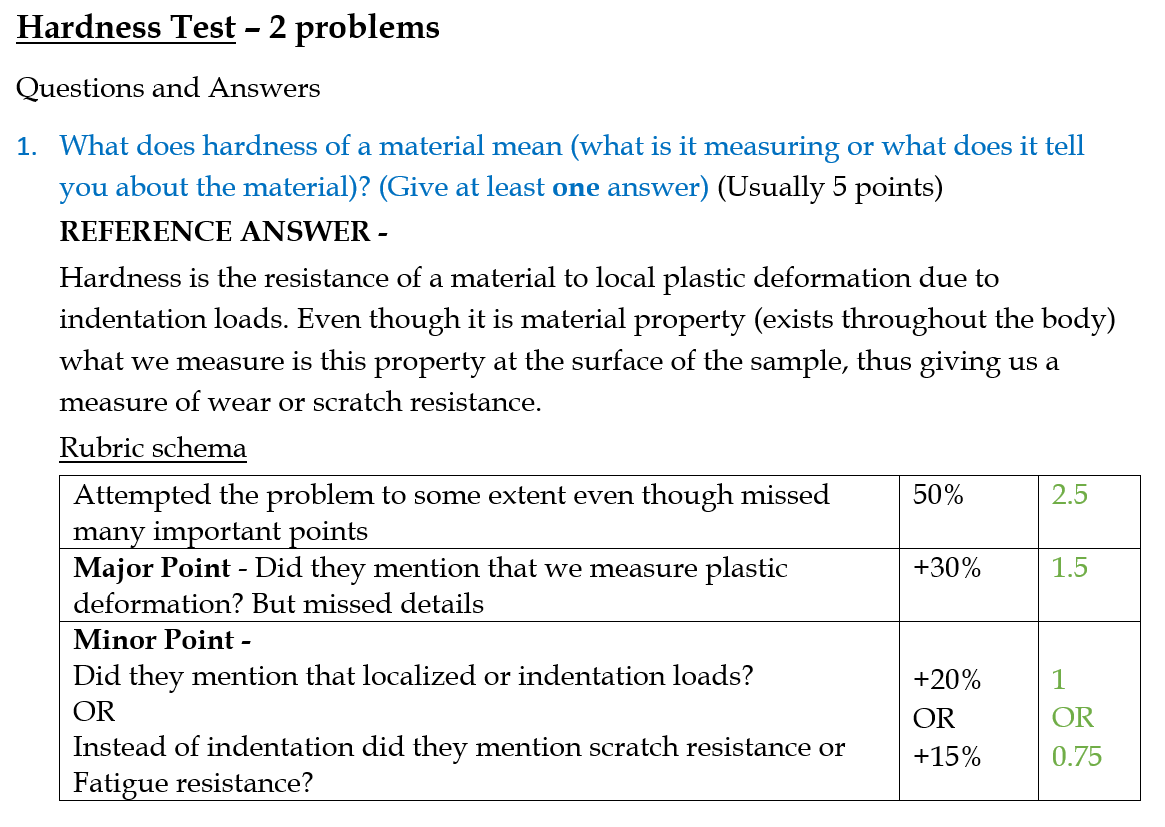}
    \caption{Hardness Test Question Example} 
    \label{fig:Hardness_Q2} 
\end{figure}

The grading problems are sourced from the Mechanical Engineering undergraduate course (MEEN 361: Materials and Manufacturing in Design Laboratory) at Texas A\&M University. The course covers material and manufacturing concepts and related lab experiments, including hardness testing, bending experiments, polymer tensile tests, fatigue testing, cold working and annealing experiments, and charpy impact testing, among others. This is a 1-credit lab course spanning 13-week Fall/Spring semester. After completing weekly lab experiments, students are required to answer 2-4 conceptual questions as a quiz, and their quiz scores contribute to their overall course grade. An example of a quiz problem and the corresponding grading rubric is shown in the \figureref{fig:Hardness_Q2}. 

\subsection{Datasets}
For this study, we selected ten quiz problems from the MEEN 361 course to conduct grading experiments (problem descriptions are provided in \appendixref{appendix_a}). Each question was answered by 225–230 students, with responses typically ranging from 3 to 7 sentences. The quizzes were graded independently by one teaching assistant (TA) and one additional grader, both trained by the course instructor using the instructor's scoring rubrics. The final scores, which serve as the "gold standard" or ground truth in this study, were verified by the instructor.

\section{Evaluation metrics}
Given the grading scale ranges from 0 to 5–10 points with several discrete scoring levels, we selected Spearman's rank correlation coefficient and Root Mean Square Error (RMSE) as the primary evaluation metrics for this study. In an educational context, rather than focusing solely on the exact score, we aim to investigate whether the machine grading can capture the same scoring trends as a human grader. To achieve this, we use Spearman's rank correlation coefficient to compare the ranking of students under the two grading methods. RMSE, on the other hand, quantifies the score differences between machine and human grading and serves as a standard metric for evaluating multi-class classification tasks.

\subsection{Spearman's Rank Correlation Coefficient}
Spearman's rank correlation coefficient assesses the extent of a monotonic relationship between student rankings under human grading and machine grading. Because graders tend to vary in the exact points awarded to the same answer, comparing raw scores directly is challenging. However, by assuming that a student's rank (i.e., relative position within the class, such as 1st, 2nd, 3rd, etc.) reflects a comparable level of knowledge across different graders, Spearman's rank correlation reduces the impact of these scoring variations. This provides a measure of consistency in the grading pattern. Spearman's rank correlation coefficient, also known as Spearman's $\rho$, is defined as follows:

\begin{equation}\label{eq:spearman_formula}
    \rho = \frac{6 \Sigma d_i^2}{n(n-1)}
\end{equation}

where $\rho$ represents the Spearman rank correlation coefficient, $d_i$ is the difference between the two ranks of each observation, and $n$ is the number of observations. We adopt this metric to evaluate whether there is a consistent monotonic relationship between the grades assigned by the TAs and those generated by GPT-4o. If GPT-4o grading achieves perfect consistency with TA grading, the ranks would exhibit a perfect linear relationship ($\rho = 1$)~\citep{gravetter2017statistics}.

Since Spearman's correlation uses ranks, it does not require dataset normalization, thereby avoiding the influence of differing score ranges across questions. 
\tableref{tab:spearman_corr} provides interpretation and strength reference ranges for both Spearman and Pearson correlation coefficients.

\begin{table}[hbtp]
\floatconts
  {tab:spearman_corr}
  {\caption{Correlation Coefficient Strengths~\citep{putnikovic2023embeddings}}}
  {\begin{tabular}{ll}
  \toprule
  \bfseries Range & \bfseries Strengths\\
  \midrule
    0.00 -- 0.19 & Very Weak \\
    0.20 -- 0.39 & Weak \\
    0.40 -- 0.59 & Moderate \\
    0.60 -- 0.79 & Strong \\
    0.80 -- 1.00 & Very Strong \\ 
  \bottomrule
  \end{tabular}}
\end{table}

\subsection{Root Mean Square Error (RMSE)}
In this experiment, RMSE is used to quantify the difference between LLM grading and TA grading (considered the gold standard). To ensure a fair comparison, datasets with varying score scales are normalized before calculating RMSE.

\begin{equation}\label{eq:spearman_formula}
    RMSE = \sqrt{\frac{\Sigma_{i=1}^{n} (y_i - \hat{y_i})^2}{n}}
\end{equation}

where $y_i$ represents the actual values, $\hat{y_i}$ denotes the predicted values, and n is the number of data points~\citep{bishop2006pattern,kuhn2013applied}. The explanation of RMSE calculations on normalized datasets is shown in \tableref{tab:rmse}.

\begin{table}[hbtp]
\floatconts
  {tab:rmse}
  {\caption{RMSE interpretation for normalized data [0, 1]}}
  {\begin{tabular}{ll}
  \toprule
  \bfseries Range & \bfseries RMSE\\
  \midrule
    0.00 -- 0.05 & Very small error \\
    0.05 -- 0.10 & Small error \\
    0.10 -- 0.20 & Moderate error \\
    0.20 -- 0.30 & Large error \\
    $ \geq 0.30 $ & Very large error \\ 
  \bottomrule
  \end{tabular}}
\end{table}


\section{Results}
The experiment was conducted using both zero-shot and few-shot prompt engineering approaches, with Spearman's $\rho$ and RMSE calculated for each. In the zero-shot prompt, no examples of student answers or grades were included, while the few-shot prompt provided a small set of student responses and corresponding grades to guide the model.

\subsection{Zero-Shot Grading with GPT-4o}

Overall, across the 10 datasets, the highest Spearman's $\rho$ observed is 0.9387, and the lowest RMSE is 0.0830, while the lowest Spearman’s rank correlation coefficient is 0.5488 and the highest RMSE is 0.2264. 

\tableref{tab:zero_shot_results} presents the Spearman's $\rho$ and RMSE results. For Spearman’s rank correlation, seven out of ten datasets exhibit coefficients over 0.6, with six datasets exceeding 0.75 and three exceeding 0.8. All datasets achieve at least 0.54, indicating that more than half of the questions (7 out of 10) demonstrate a strong and above correlation between GPT-4o automated grading and TA grading. Among these, three datasets show a very strong correlation, and all datasets achieve at least a moderate correlation.

For RMSE, three out of ten datasets exhibit small errors (0.05 – 0.10), indicating that GPT-4o grading closely aligns with TA grading in these cases. The highest RMSE observed is 0.2264, with 9 datasets recording RMSE values below 0.2 and 1 dataset (Fatigue 2) mildly above 0.2, signifying mostly a moderate error level. This suggests that while GPT-4o grading is generally consistent with TA grading, there are some discrepancies. 

\begin{table}[hbtp]
\floatconts
  {tab:zero_shot_results}
  {\caption{Zero-shot results}}
  {\begin{tabular}{lcc}
  \toprule
  \bfseries Question Name & \bfseries Spearman’s $\rho$ & \bfseries RMSE\\
  \midrule
    Charpy Impact Q1 & \textbf{0.6601} & 0.1566 \\
    Cold Working and Annealing Q1 & \textbf{\underline{0.9387}} & \textbf{0.0975} \\
    Fatigue Q2 & \textbf{0.7694} & \underline{0.2264} \\
    Hardness Q1 & \textbf{\underline{0.8183}} & \textbf{0.0830} \\
    Three-Point Bending Q1 & 0.5518 & 0.1566 \\
    Three-Point Bending Q2 & 0.5488 & 0.1629 \\
    Three-Point Bending Q4 & \textbf{0.7524} & 0.1758 \\
    Polymer Tensile Test Q1 & \textbf{0.7574} & 0.1622 \\
    Polymer Tensile Test Q2 & 0.5850 & 0.1819 \\
    Polymer Tensile Test Q4 & \textbf{\underline{0.8911}} & \textbf{0.0872} \\
  \bottomrule
  \end{tabular}}
\end{table}

\figureref{fig:zero-shot-best} and \figureref{fig:zero-shot-worst} illustrate the LLM and TA grades for the datasets with the best and worst Spearman's $\rho$ values, respectively. In these figures, the blue line represents TA grading, while the red line represents LLM grading.

\begin{figure}[htbp]
\floatconts
  {fig:zero-shot-best}
  {\caption{Cold Working and Annealing Quiz Q1 grading results (best 0.9387)}}
  {\includegraphics[width=0.9\linewidth]{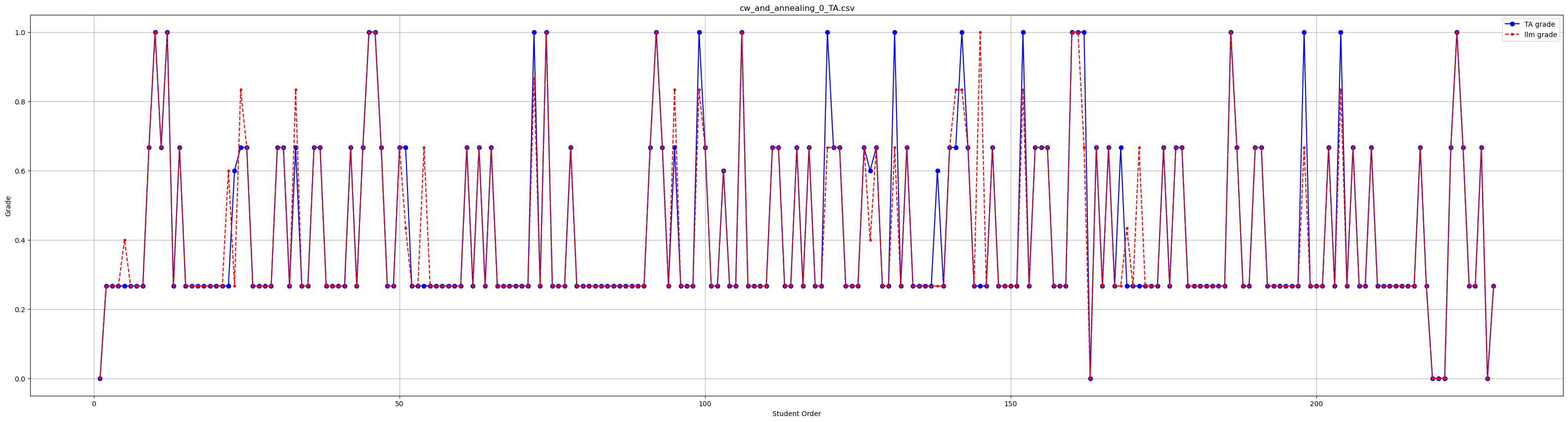}}
\end{figure}

\begin{figure}[htbp]
\floatconts
  {fig:zero-shot-worst}
  {\caption{Three-Point Bending Q2 grading results (worst 0.5488)}}
  {\includegraphics[width=0.9\linewidth]{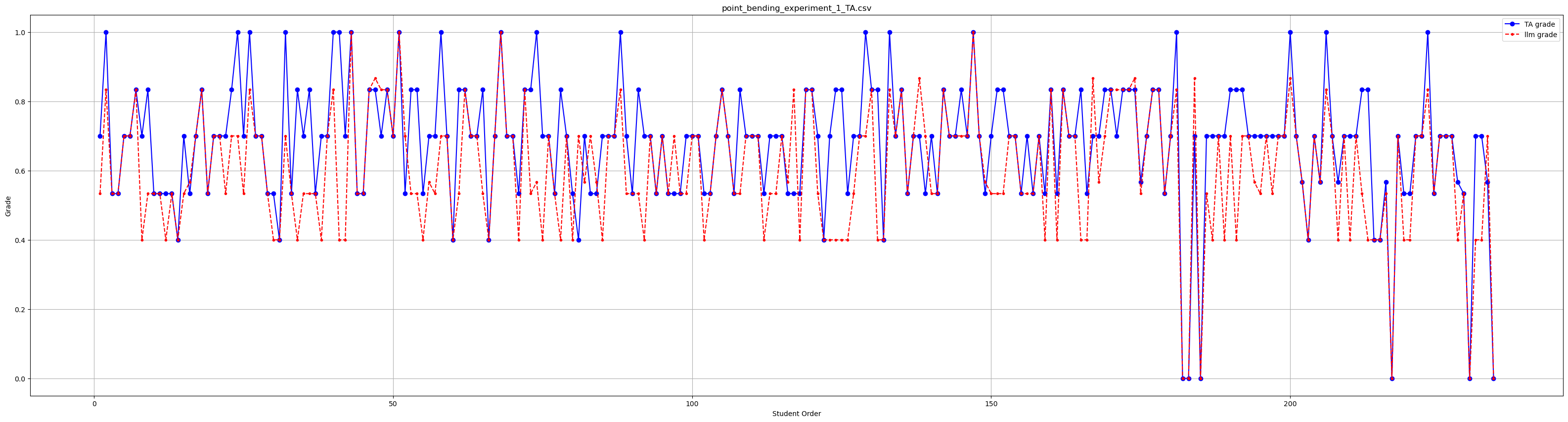}}
\end{figure}

\subsection{In-Context Learning Approach: Few-Shot Grading with GPT-4o}

As a comparison, we experiment with in-context learning strategy by providing relevant TA grading examples to GPT-4o to investigate the model's performance.

In the few-shot prompt setup, the highest Spearman’s rank correlation coefficient achieved is 0.8990, with the lowest RMSE at 0.0998. The lowest Spearman’s rank correlation coefficient observed is 0.5202, and the highest RMSE is 0.3733. Following the initial zero-shot prompt run, we selected four student answers and their corresponding grades—where there was an agreement between TA and LLM grading—from high to low scores as examples to include in the prompt.

\tableref{tab:few_shot_results} shows Spearman’s rank correlation coefficient and RMSE results. For  Spearman's $\rho$, seven out of ten datasets show coefficients above 0.6. However, values remain largely unchanged or decrease for most datasets. Specifically, three datasets exceed 0.75, which is three fewer than in the zero-shot setup, and only one dataset exceeds 0.8. All datasets record coefficients above 0.52, indicating that, although adding examples slightly decreased the correlation between GPT-4o grading and TA grading, all datasets still exhibit at least a moderate correlation.

In terms of RMSE, only one out of nine datasets shows a small error (0.05–0.10), a reduction from the three datasets observed in the zero-shot configuration. However, except Fatigue 2 RMSE, other 9 datasets still maintain RMSE values below 0.2, indicating a moderate level of error.

\begin{table}[hbtp]
\floatconts
  {tab:few_shot_results}
  {\caption{Few-shot results}}
  {\begin{tabular}{lcc}
  \toprule
  \bfseries Question Name & \bfseries Spearman’s $\rho$ & \bfseries RMSE\\
  \midrule
         Charpy Impact Quiz Q1 &
0.5585&
0.1697 \\
Cold Working and Annealing Quiz Q1 &
\textbf{\underline{0.8990}} &
0.1238\\

Fatigue Q2 &
\textbf{0.6913} &
\underline{0.3733}\\

Hardness Quiz Q1 &
\textbf{0.7069} &
\textbf{0.0998}\\

Three-Point Bending Q1 &
0.5531 &
0.1508\\

Three-Point Bending Q2 &
0.5202&
0.1661\\

Three-Point Bending Q4 &
\textbf{0.7564} &
0.1596\\

Polymer Tensile Test Quiz Q1 &
\textbf{0.6880} &
0.1815 \\

Polymer Tensile Test Quiz Q2 &
\textbf{0.6120} &
0.1579 \\
Polymer Tensile Test Quiz Q4 &
\textbf{0.7638} &
0.1308\\

  \bottomrule
  \end{tabular}}
\end{table}

\figureref{fig:few-shot_cw} and \figureref{fig:few-shot-worst} display the LLM and TA grades for the datasets with the highest and lowest Spearman’s $\rho$, respectively. In these figures, the blue line represents TA grading, while the red line represents LLM grading.

\begin{figure}[htbp]
\floatconts
  {fig:few-shot_cw}
  {\caption{Cold Working and Annealing Quiz Q1 grading results (best 0.8990)}}
  {\includegraphics[width=0.9\linewidth]{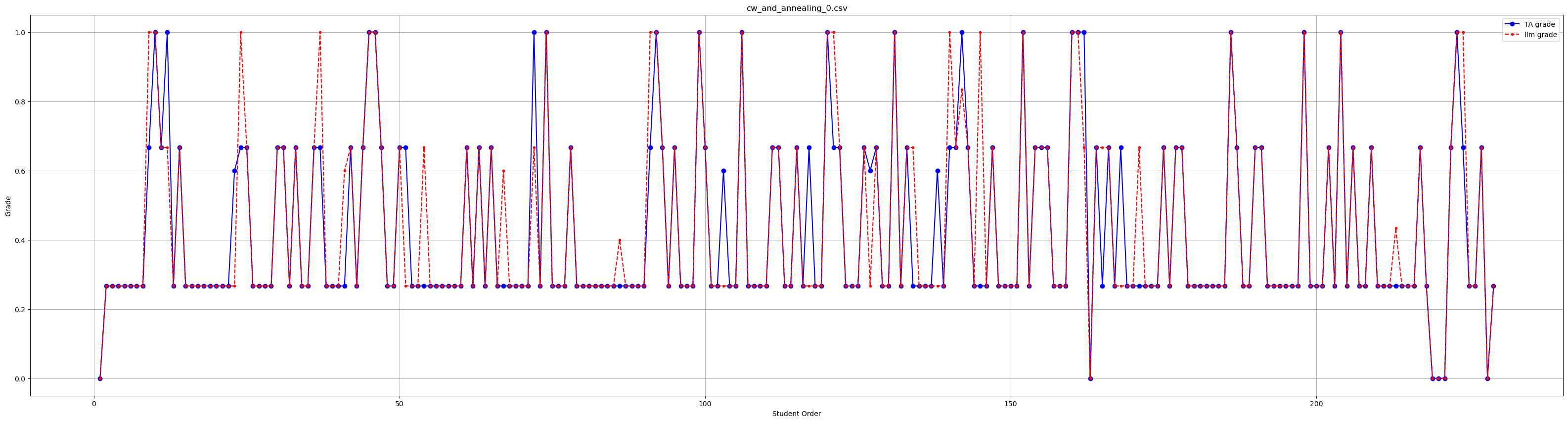}}
\end{figure}

\begin{figure}[htbp]
\floatconts
  {fig:few-shot-worst}
  {\caption{Three-Point Bending Q2 grading results (worst 0.5202)}}
  {\includegraphics[width=0.9\linewidth]{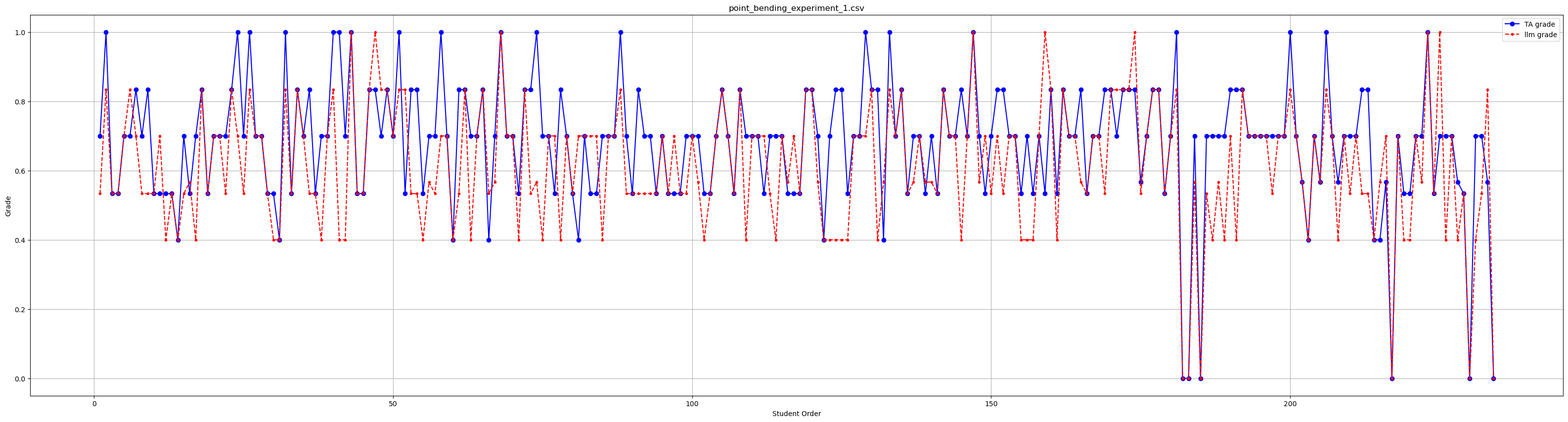}}
\end{figure}

\subsection{Comparison of In-Context learning}
A comparison of the results with/without in-context learning is provided in \tableref{tab:shot_compare}. Generally, datasets with an increased Spearman’s $\rho$ also exhibit a decrease in RMSE. However, only 3 out of 10 datasets show mild improvement, while the performance in 7 out of 10 datasets declines. This indicates that adding four graded examples did not significantly enhance the model's performance in grading short conceptual questions in mechanical engineering. Although some datasets showed slight improvement, it was minimal. Potential reasons for this outcome include: 

\begin{itemize}
    \item Student responses to open-ended conceptual questions tend to vary widely, making it likely that the examples provided led the LLM in an unintended direction.
    
    \item With a dataset size of approximately 225 to 230, the addition of four examples may have been excessive, potentially leading to overfitting.\footnote{We opted not to use random training and testing datasets, as this approach would be impractical. If the TA must grade a large number of questions to train the LLM, it may undermine the purpose of automating the grading process.}
    
    \item Furthermore, given the presence of a reference (or "golden") answer, additional examples may offer limited benefit, suggesting that adding examples could be counterproductive.
\end{itemize}

\begin{table}[hbtp]
\floatconts
  {tab:shot_compare}
  {\caption{Comparison of few-shot and zero-shot prompt results}}
  {\begin{tabular}{lcc}
  \toprule
  \bfseries Question Name & \bfseries $\boldsymbol{\rho}_{\boldsymbol{few}} - \boldsymbol{\rho}_{\boldsymbol{zero}}$ & \bfseries $\boldsymbol{RMSE}_{\boldsymbol{few}} - \boldsymbol{RMSE}_{\boldsymbol{zero}}$\\
  \midrule
         Charpy Impact Quiz Q1 &
-0.1016 &
0.0131\\
Cold Working and Annealing Quiz Q1 &
-0.0397 &
0.0263\\

Fatigue Q2 &
-0.0781 &
0.1469\\

Hardness Quiz Q1 &
-0.1114 &
0.0168\\

Three-Point Bending Q1 &
\textbf{0.0013}&
\textbf{-0.0058}\\

Three-Point Bending Q2 &
-0.0286 &
0.0032\\

Three-Point Bending Q4 &
\textbf{0.0040} &
\textbf{-0.0162} \\

Polymer Tensile Test Quiz Q1 &
-0.0694 &
0.0193 \\

Polymer Tensile Test Quiz Q2 &
\textbf{0.0270} &
\textbf{-0.0240} \\
Polymer Tensile Test Quiz Q4 &
-0.1273 &
0.0436\\

  \bottomrule
  \end{tabular}}
\end{table}

\section{Grading Performance Analysis}

Several key observations emerged when comparing GPT-4o's grading performance with TA grading:

\begin{enumerate}
    \item \textbf{Strong Performance on Clear Scoring Rubrics:}  GPT-4o model exhibited high accuracy when the scoring criteria were straightforward and clearly defined. However, its performance declined with more complex or nuanced questions, where precise interpretation was required.
    \item \textbf{Difficulty with Synonyms:} GPT-4o struggled with responses that used synonymous terms not explicitly covered by the rubric. In such cases, the model was likely to assign lower scores despite the semantic correctness of the answers.
    \item \textbf{Stricter Scoring on Ambiguity:} In scenarios where the scoring criteria were unclear or ambiguous, GPT-4o generally applied more stringent grading compared to human TAs, who tended to give higher scores in similar situations.
    \item \textbf{High Accuracy on High-Weight Items:} The model performed well in cases where specific answer choices carried significant weight in the rubric, accurately identifying and scoring these key elements.
\end{enumerate}

In comparing human and LLM-based grading, we observe that LLM grading shows promise in evaluating short answer questions. In many cases, GPT-4o successfully assessed students' responses according to the rubric criteria provided. One notable observation was related to a rubric criterion focused on lab observation, which awards a minimum score to students who made an attempt on the question. We found that the LLM's grading of this particular rubric point remained consistent across responses without any misinterpretation. However, the LLM's grading of other rubric points was more variable, displaying a mix of accurate and less precise assessments.

In several instances, the grades assigned by the LLM aligned closely with those given by human evaluators, demonstrating its potential for effective grading. Nevertheless, there were notable instances where the LLM missed key details. For example, when students mentioned the “DBTT” (Ductile to Brittle Transition Temperature), the GPT-4o correctly identified and credited this point. However, it sometimes failed to recognize significant information when students explained that steel becomes brittle and breaks easily at cold temperatures. A similar issue occurred with the concept of ultimate strength: while GPT-4o reliably awarded points when students mentioned it, it occasionally granted points even when students did not reference it.

\section{Confusion Matrix}
This inconsistency underscores the need for instructors to provide clearer contextual guidance. Testing a small set of sample answers may help refine the LLM’s ability to address these gaps effectively. As a grading assistant, this tool could offer instructors insights into answer patterns, supporting the grading process.

We selected the Charpy Impact Test Q1 as an example for confusion matrix analysis, shown in \figureref{fig:confusion-matrix}. Most of the machine-graded scores aligned with human grading; however, in the inconsistent cases, a trend emerges where the machine scores are generally lower than human scores. It is also important to note that the true labels in this study are based on human grading, which can also include inherent biases since different TAs have subjective biases.

\begin{figure}[htbp]
\floatconts
  {fig:confusion-matrix}
  {\caption{Charpy Impact Test Q1 - Confusion Matrix}}
  {\includegraphics[width=0.6\linewidth]{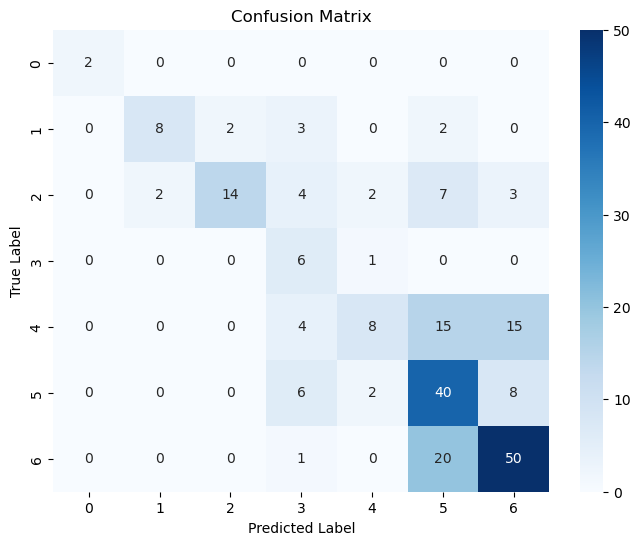}}
\end{figure}

\section{Conclusion and Future Work}

This study demonstrates the potential of using large language models (LLMs), specifically GPT-4o, for automated grading of conceptual questions in an undergraduate Mechanical Engineering course. In the zero-shot prompt setting, GPT-4o achieved a strong correlation with human teaching assistants (TAs), particularly when grading tasks were aligned with clear and straightforward rubrics. The model's performance remained consistent in recognizing key concepts, but it exhibited limitations in handling nuanced answers, especially when synonyms or ambiguous terms not explicitly addressed in the rubric were used. Moreover, the in-context learning approach, intended to improve performance by providing example answers, did not consistently enhance the model's accuracy and, in some cases, introduced variability in the grading outcomes.

Despite these challenges, GPT-4o holds significant promise as a scalable tool for grading in large classroom settings. It offers the potential to reduce the grading workload for educators while maintaining overall consistency with human grading. However, the model's tendency to grade more stringently and its struggles with less explicit criteria highlight the need for further refinements in rubric design and prompt engineering.

Future work will focus on addressing the limitations identified in this study. Specifically, efforts will be made to:
\renewcommand{\labelenumi}{\theenumi)} 
\begin{enumerate}
    \item Refine Rubric and Prompt Design: Enhancing the clarity and specificity of rubrics and improving prompt engineering strategies to better guide ChatGPT model in interpreting nuanced and diverse student responses. This includes exploring ways to balance the number of few-shot examples provided to avoid overfitting or introducing noise.
    \item Fine-Tuning LLMs for Domain-Specific Grading: Investigating the fine-tuning of open-source models on domain-specific datasets, particularly those related to Mechanical Engineering, to improve the open-source model's understanding of technical terminology and common student answer patterns.
    \item Expand Dataset and Problem Types: Extending the study to include a broader range of conceptual questions and larger datasets to assess the generalizability of the findings across different types of conceptual questions and STEM disciplines.
\end{enumerate}

By addressing these areas, future work aims to enhance the accuracy, scalability, and practical application of ChatGPT and similar LLMs in educational assessment.

\acks{We gratefully acknowledge Jiachang Xing for her assistance with dataset grading in this study.}

\bibliography{pmlr-sample}

\clearpage
\appendix
\section{Conceptual Question Description and Grading Rubric}\label{appendix_a}

\begin{figure}[htbp]
\floatconts
    {fig:appendix-1}
    {\caption{Dataset 1: Charpy Impact Quiz Q1}}
    {\includegraphics[width=0.6\linewidth]{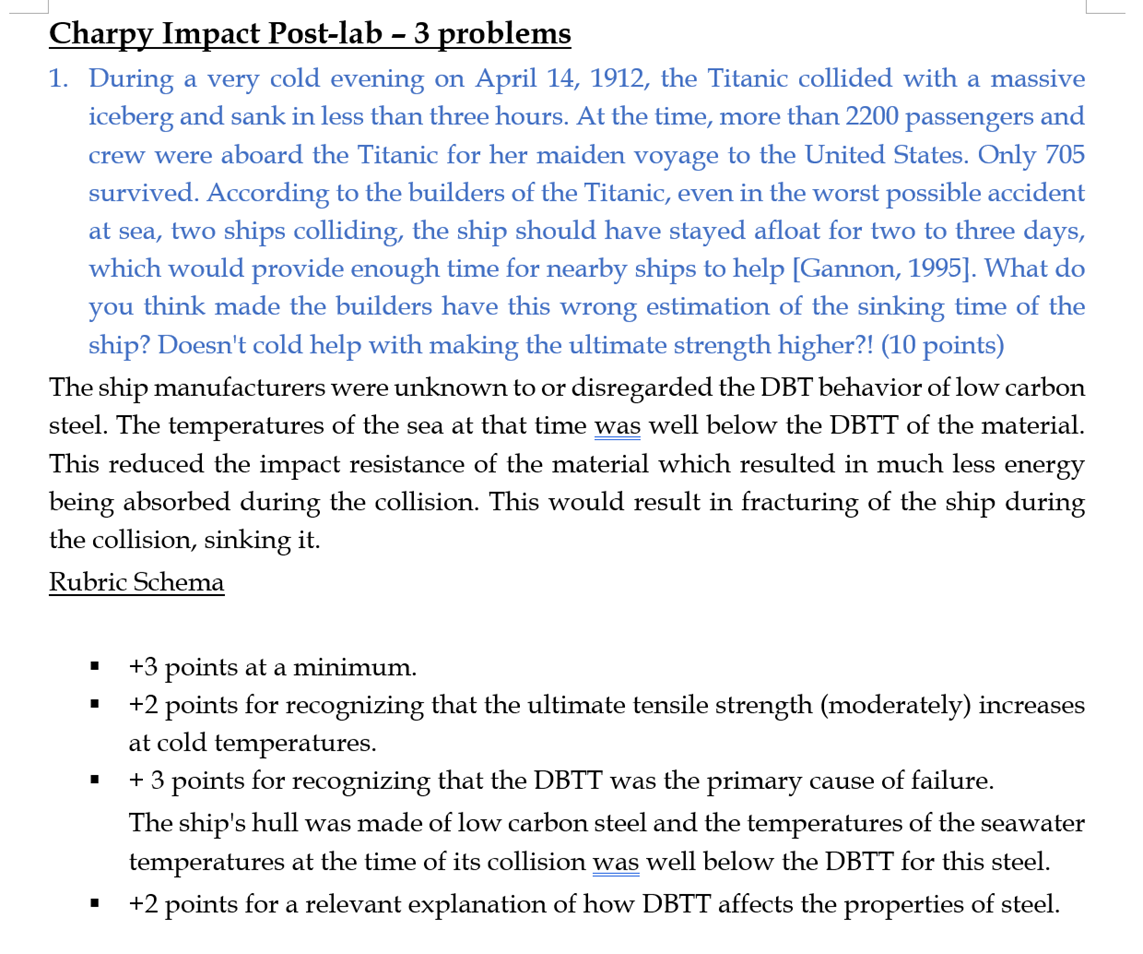}}
\end{figure}

\begin{figure}[htbp]
\floatconts
    {fig:appendix-2}
    {\caption{Dataset 2: Cold Working and Annealing Quiz Q1}}
    {\includegraphics[width=0.6\linewidth]{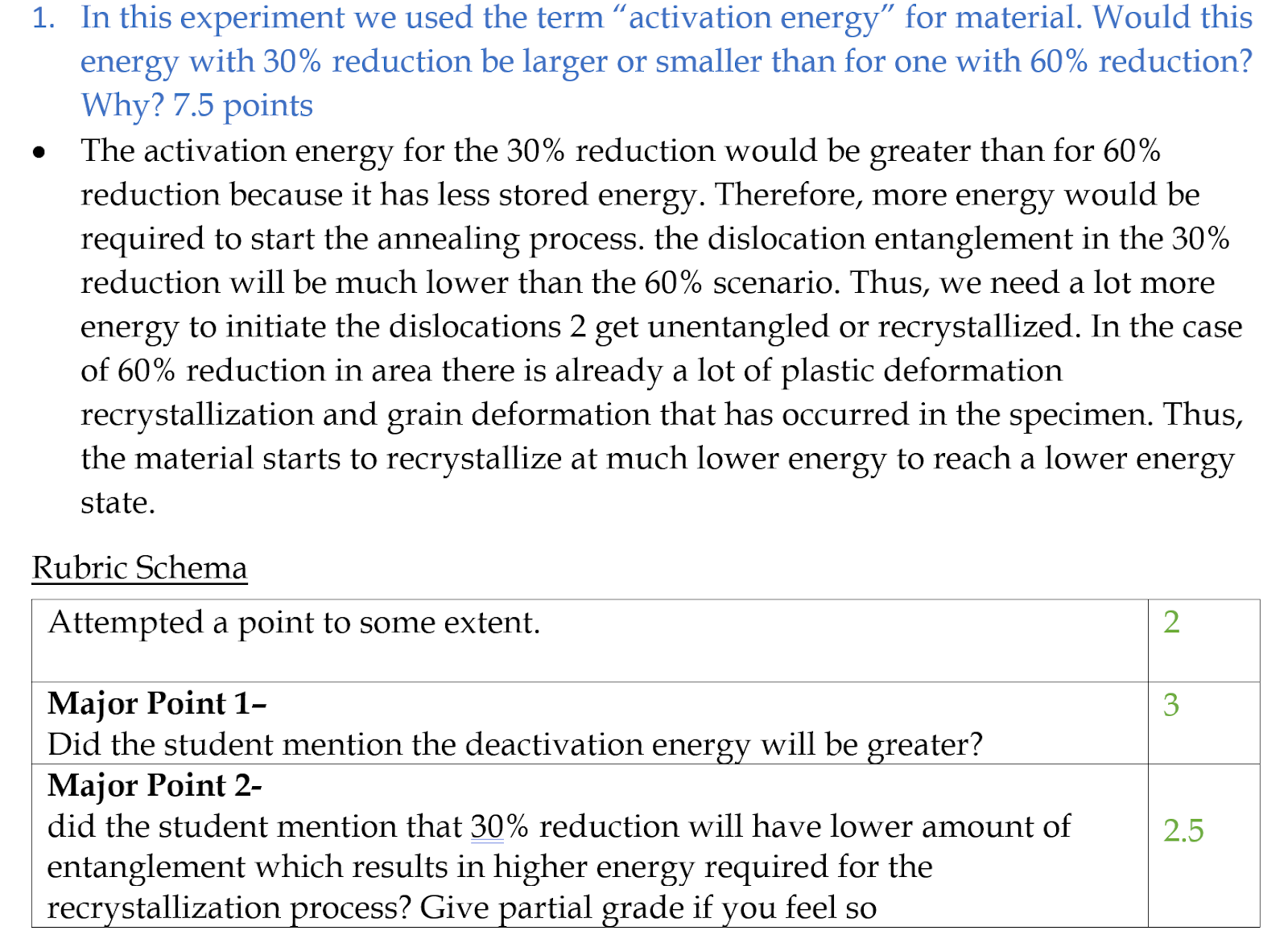}}
\end{figure}

\begin{figure}[htbp]
\floatconts
    {fig:appendix-3}
    {\caption{Dataset 3: Fatique Quiz Q2}}
    {\includegraphics[width=0.6\linewidth]{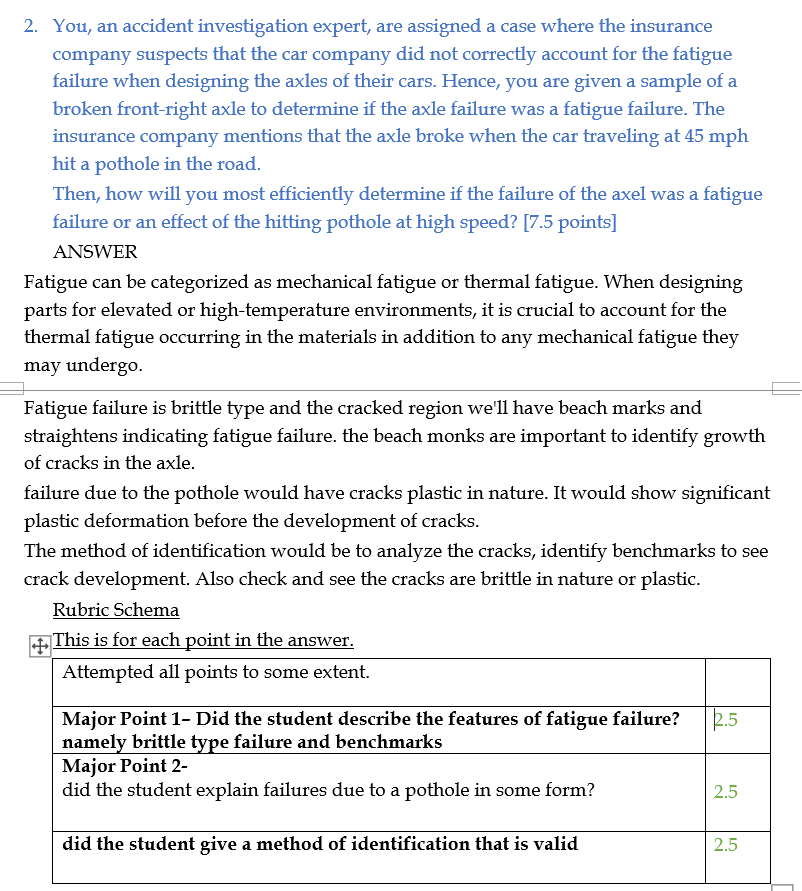}}
\end{figure}

\begin{figure}[htbp]
\floatconts
    {fig:appendix-3}
    {\caption{Dataset 4: Hardness Quiz Q1}}
    {\includegraphics[width=0.6\linewidth]{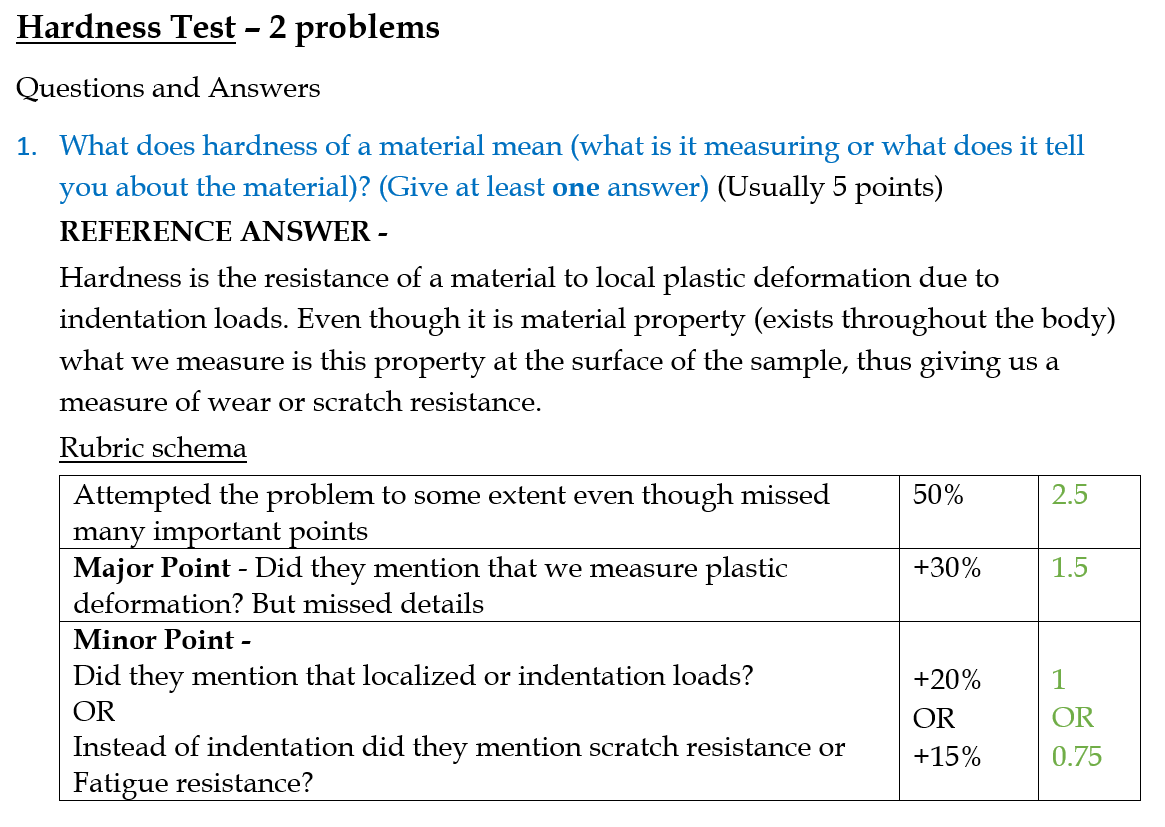}}
\end{figure}

\begin{figure}[htbp]
\floatconts
    {fig:appendix-4}
    {\caption{Dataset 5: Three-Point Bending Q1}}
    {\includegraphics[width=0.6\linewidth]{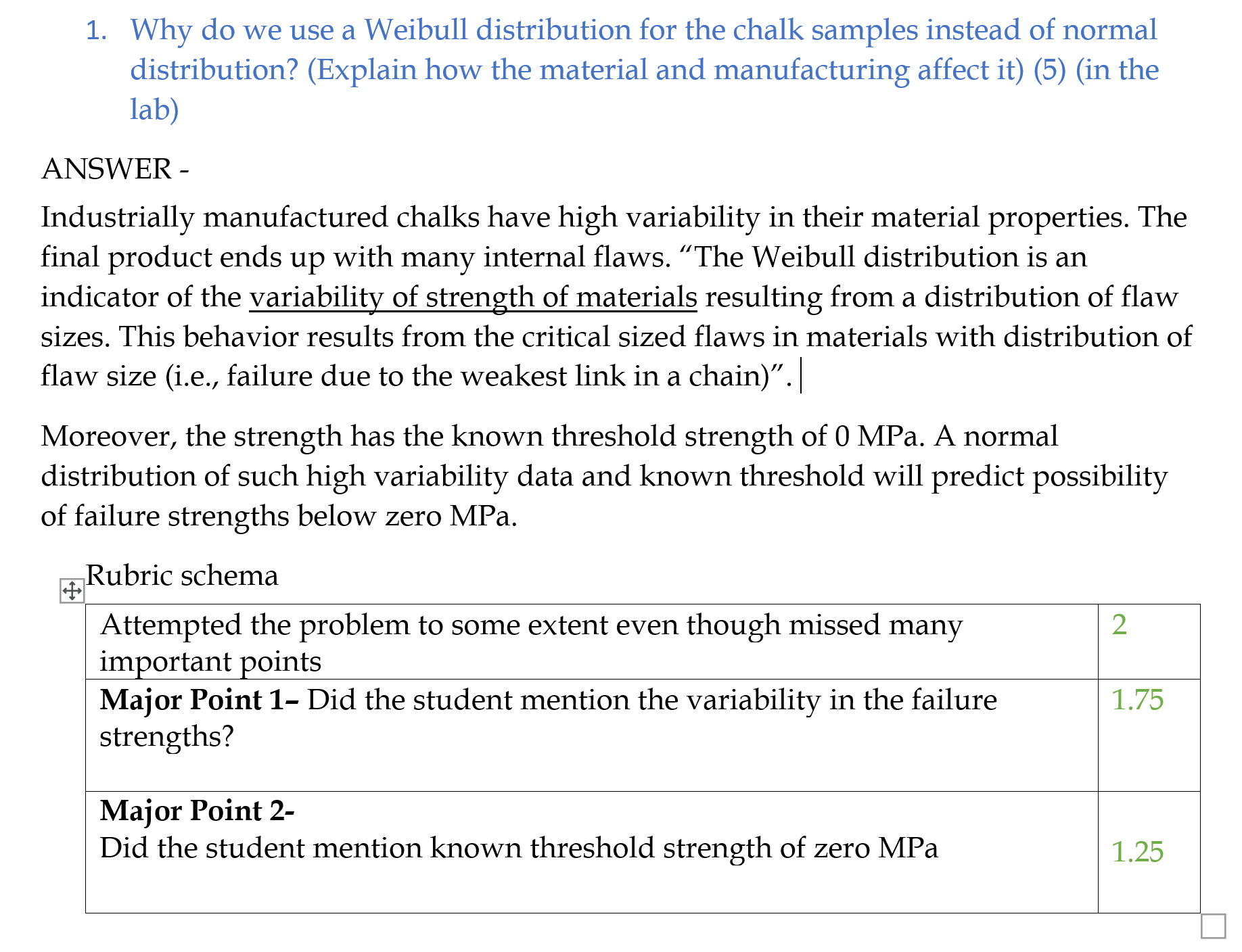}}
\end{figure}

\begin{figure}[htbp]
\floatconts
    {fig:appendix-5}
    {\caption{Dataset 6: Three-Point Bending Q2}}
    {\includegraphics[width=0.6\linewidth]{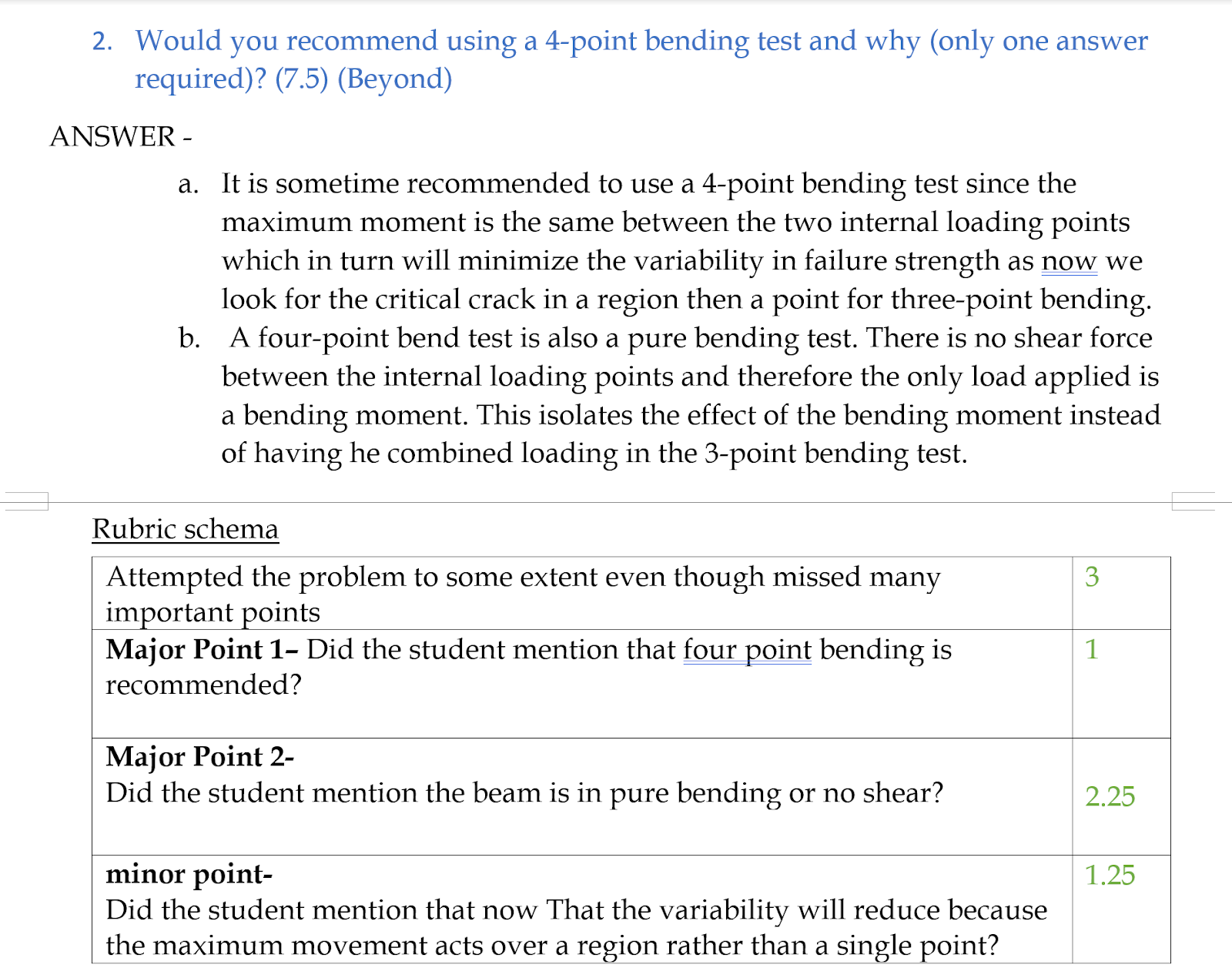}}
\end{figure}

\begin{figure}[htbp]
\floatconts
    {fig:appendix-6}
    {\caption{Dataset 7: Three-Point Bending Q4}}
    {\includegraphics[width=0.6\linewidth]{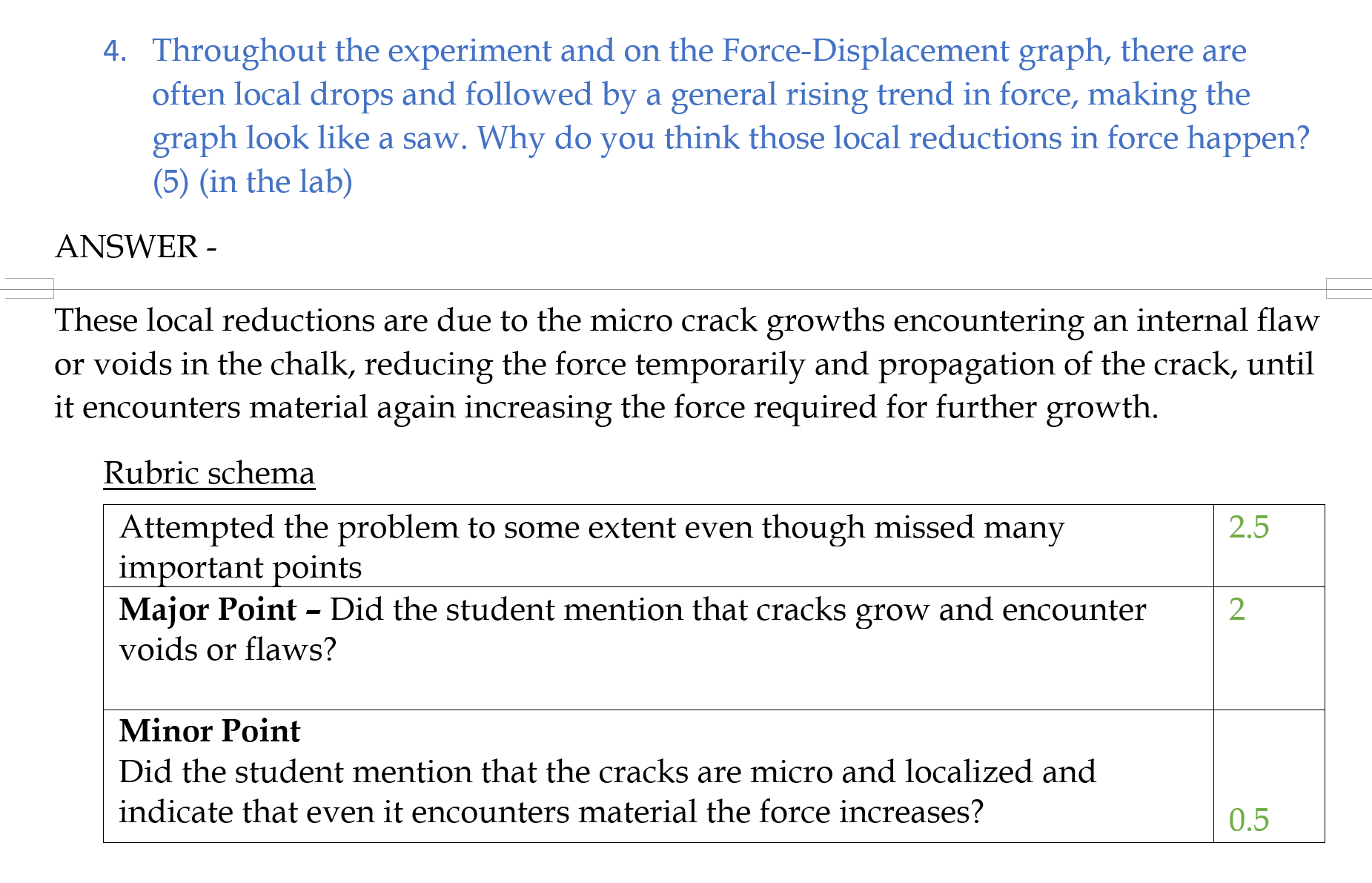}}
\end{figure}

\begin{figure}[htbp]
\floatconts
    {fig:appendix-7-1}
    {\caption{Dataset 8: Polymer Tensile Test Quiz Q1}}
    {\includegraphics[width=0.6\linewidth]{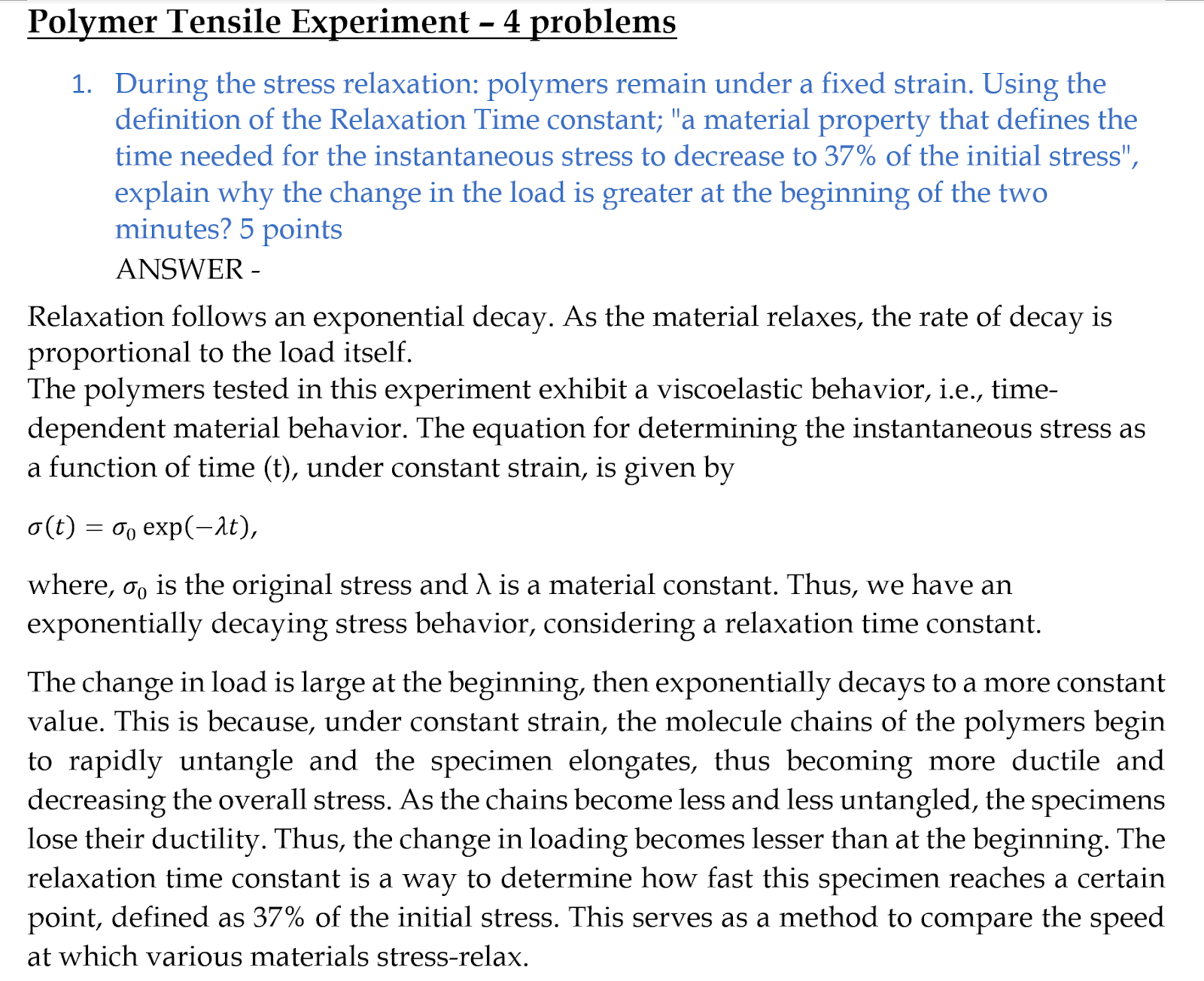}}
\end{figure}
\begin{figure}[htbp]
\floatconts
    {fig:appendix-7-2}
    {\caption{Dataset 8: Polymer Tensile Q1 Rubric}}
    {\includegraphics[width=0.6\linewidth]{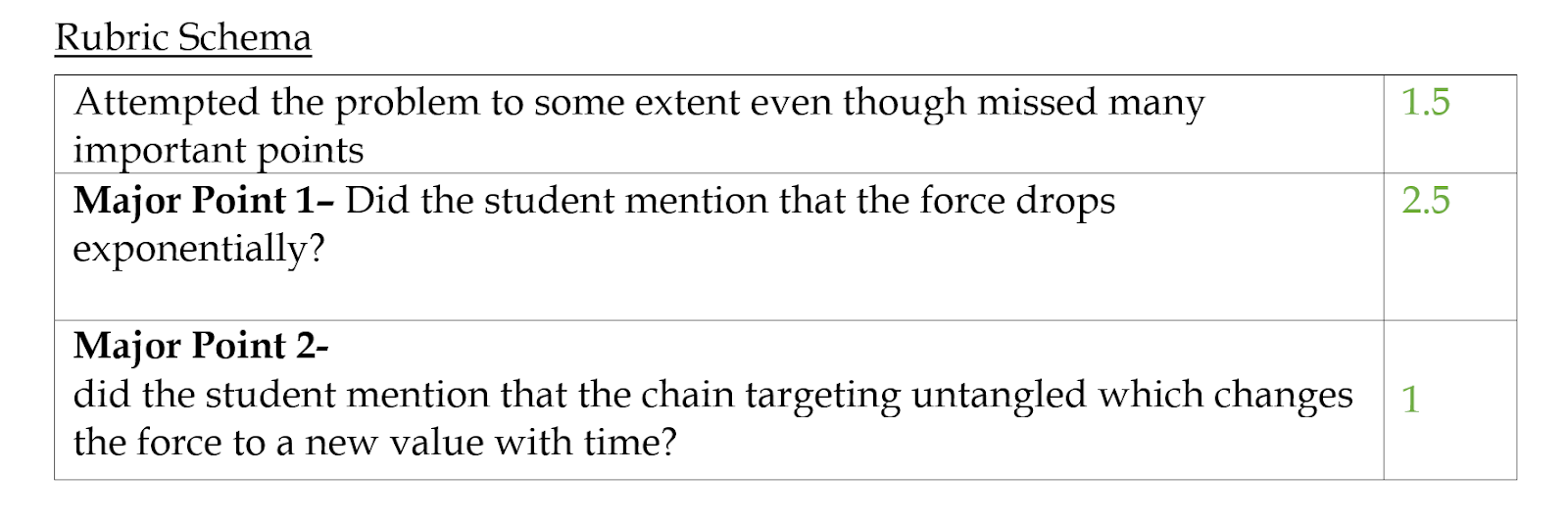}}
\end{figure}

\begin{figure}[htbp]
\floatconts
    {fig:appendix-8-1}
    {\caption{Dataset 9: Polymer Tensile Test Quiz Q2}}
    {\includegraphics[width=0.6\linewidth]{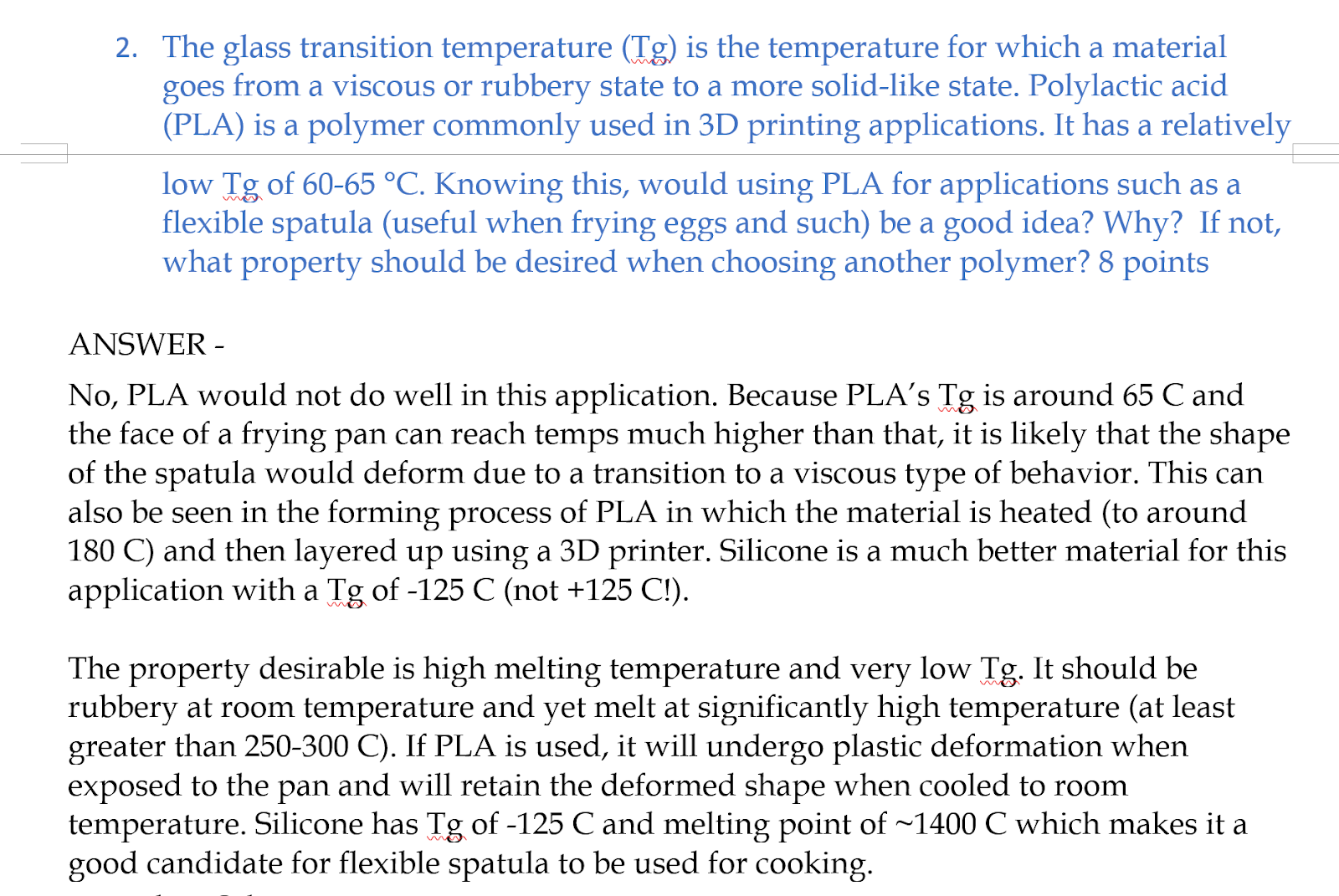}}
\end{figure}

\begin{figure}[htbp]
\floatconts
    {fig:appendix-8-2}
    {\caption{Dataset 9: Polymer Tensile Q1 Rubric}}
    {\includegraphics[width=0.6\linewidth]{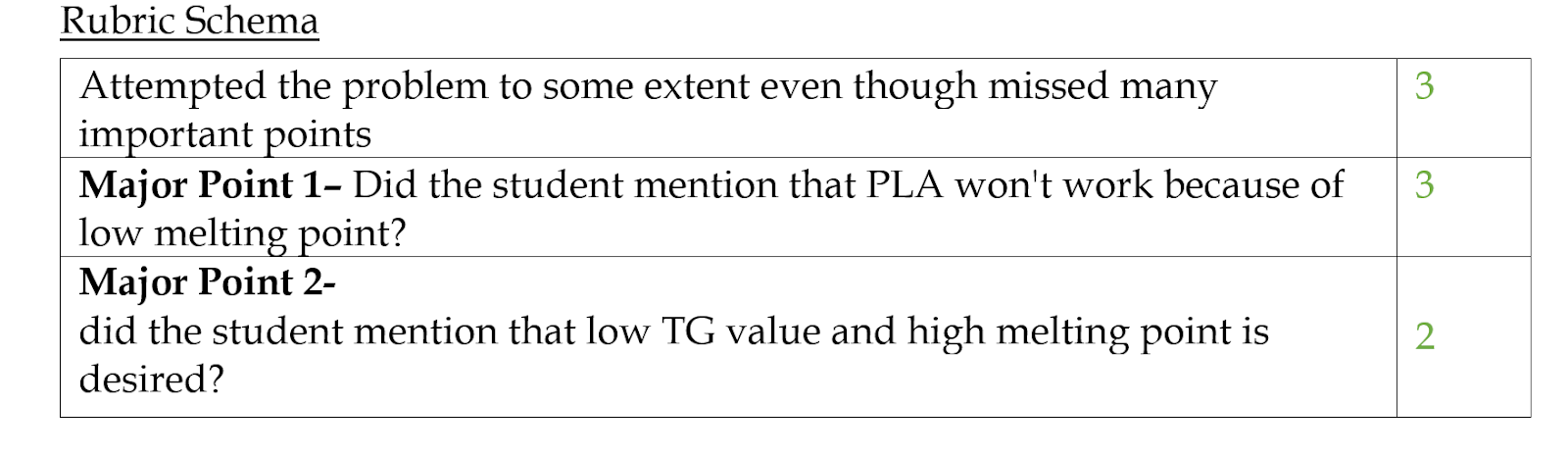}}
\end{figure}

\begin{figure}[htbp]
\floatconts
    {fig:appendix-9}
    {\caption{Dataset 10: Polymer Tensile Test Quiz Q4}}
    {\includegraphics[width=0.6\linewidth]{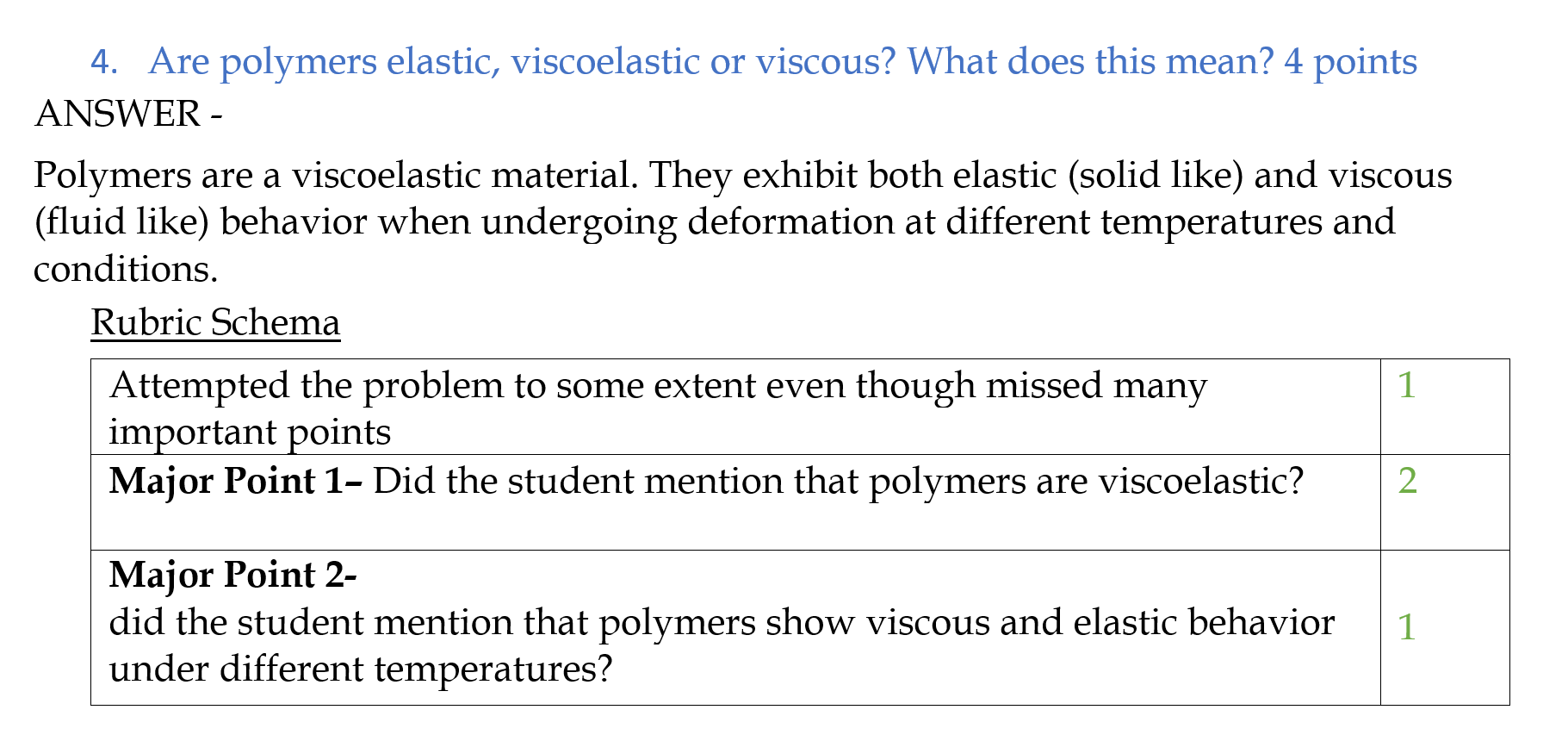}}
\end{figure}

\clearpage
\section{Prompt example}\label{appendix_b}

\begin{figure}[htbp]
\floatconts
    {fig:appendix-10}
    {\caption{Zero-shot grading experiment prompt}}
    {\includegraphics[width=0.9\linewidth]{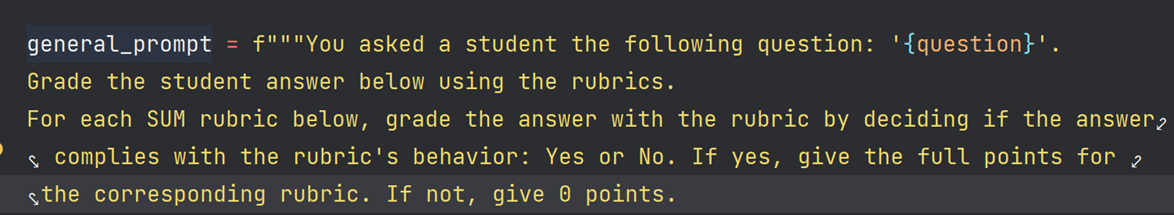}}
\end{figure}

\begin{figure}[htbp]
\floatconts
    {fig:appendix-11}
    {\caption{Few-shot grading experiment prompt}}
    {\includegraphics[width=0.9\linewidth]{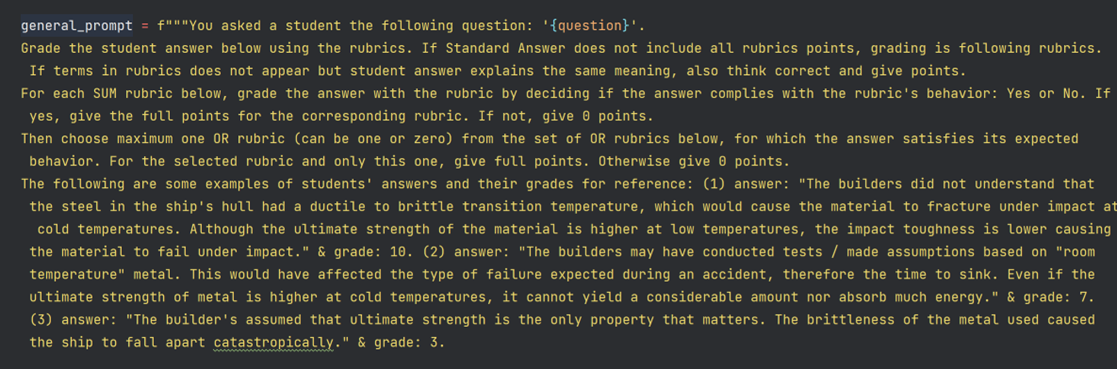}}
\end{figure}

\end{document}